# Enhanced Light Extraction and Beam Focusing in GaN LEDs Using Hybrid Metasurface-Distributed Bragg Reflector Structures

Hanbo Xu[1], Xinyang Liu[2], Lei Wang[1,*]

[1]*College of Physics, Jilin University, Changchun, 130012, China*
[2]*School of Mechanical and Aerospace Engineering, Jilin University, Changchun, 130025, China*
*Corresponding author: Lei Wang, wang_lei@jlu.edu.cn

**Abstract**

This study presents an optimized hybrid design integrating $TiO_2$ nanocylinder metasurfaces with Distributed Bragg Reflectors (DBRs) to simultaneously enhance light extraction efficiency (LEE) and beam collimation in GaN-based light-emitting diodes (LEDs). Through systematic theoretical modeling and numerical simulations using COMSOL and Finite-Difference Time-Domain (FDTD) methods, we investigate the impact of DBR layer count and metasurface geometry—including nanocylinder radius, height, and periodicity—on optical performance. The optimized structure achieves a narrow divergence angle of 5.7° alongside an LEE of 25.67%, demonstrating competitive performance compared to existing technologies. The DBR enhances reflectivity to minimize trapped light, while the $TiO_2$ metasurface leverages Mie resonance for precise beam control. Parametric studies reveal that a single-layer DBR (46 nm $TiO_2$ / 77 nm $SiO_2$) combined with a nanocylinder metasurface (radius = 71 nm, height = 185 nm, periodicity = 222 nm) optimally balances efficiency and directionality. Far-field analysis confirms strong main lobe intensity (0.00752 V/m) with suppressed sidelobes, ensuring high-directionality emission. Compared to prior works, our design achieves superior beam collimation without significant LEE trade-offs, addressing key challenges in LED performance such as photon recycling losses and angular dispersion. These findings provide a practical framework for advancing high-brightness, directional LEDs in applications like micro-displays, LiDAR, and optical communications. Future work will explore broadband optimization and polarization control to further enhance performance.

## 1. Introduction

The growing demand for energy-efficient light sources has positioned light-emitting diodes (LEDs) as a cornerstone of modern optoelectronic technology. LEDs are widely adopted due to their superior energy efficiency, compact form factor, extended operational lifespan, and versatility across applications ranging from general lighting to advanced display technologies [1-4]. Among semiconductor-based LEDs, gallium nitride (GaN) devices have emerged as particularly promising due to their wide bandgap, high thermal stability, and ability to emit across the visible and ultraviolet spectra. However, despite significant advancements in GaN LED technology, a persistent challenge remains in maximizing light extraction efficiency (LEE). The high refractive index contrast between GaN ($n \approx 2.4$) and the surrounding medium (typically air or encapsulation materials with $n \approx 1.0–1.5$) results in a substantial portion of generated light being trapped within the LED structure due to total internal reflection (TIR). This phenomenon leads to significant optical losses, directly impairing the external quantum efficiency (EQE) and overall device performance [3, 4].

To mitigate these losses, researchers have explored multiple strategies aimed at enhancing LEE. Early approaches included surface roughening techniques, which reduce Fresnel reflections by randomizing the light escape path, and photonic crystal structures, which employ periodic dielectric patterns to manipulate light propagation via Bragg diffraction [3,5,6]. While these methods have demonstrated moderate success, they often suffer from fabrication complexity, limited spectral bandwidth, or insufficient control over emission directionality. More recently, metasurface technology has emerged as a transformative solution for light management in LEDs [3,5-7]. Metasurfaces—composed of subwavelength-scale nanostructures—enable unprecedented control over light by locally tailoring its amplitude, phase, and polarization. When integrated with LEDs, metasurfaces can significantly enhance LEE while simultaneously enabling precise beam shaping, a capability that is critical for applications requiring highly directional emission, such as micro-displays, LiDAR, and optical communications [8].

Recent studies have demonstrated the potential of metasurface-integrated GaN LEDs in improving both LEE and emission directionality. For instance, Ge et al. developed a dual-hemisphere photonic crystal structure on GaN LEDs using ZnO/GaN/SiC heterostructures, achieving an LEE of 27.16% through finite-difference time-domain (FDTD) optimization. However, this work did not address beam collimation, leaving the divergence angle unoptimized

[9]. In a bio-inspired approach, Mao et al. implemented a disordered metasurface of silver nanoparticles (Ag NPs) on GaN LEDs, boosting the EQE from 31.6% to 51.5% and LEE from 64.4% to 73.6%. While this design achieved exceptional LEE, the emitted light remained diffuse rather than directional, limiting its applicability in systems requiring focused beams [10]. Alternative strategies have employed distributed Bragg reflectors (DBRs) to enhance on-axis emission. Huang et al. fabricated GaN-based resonant cavity micro-LEDs with $SiO_2/TiO_2$ DBRs, reducing the divergence angle to 78.7° for 150 μm devices—a notable improvement over conventional micro-LEDs. However, this study did not quantify LEE, leaving the overall extraction efficiency unverified [11]. More recently, researchers have demonstrated InGaN-based blue resonant cavity micro-LEDs (RC-μ-LEDs) incorporating staggered multiple quantum wells (SMQWs) and nanoporous DBRs, achieving a divergence angle of 39.04° [12]. Another breakthrough was reported by Chen et al., who designed a broadband beam-collimating metasurface for full-color micro-LEDs, reducing divergence angles from ±65.44°–±66.87° to as low as ±3.17°–±6.64° while increasing central light intensity by up to 42-fold. Despite these advances, the study did not report explicit LEE values, leaving the trade-off between extraction efficiency and beam control unresolved [13].

While metasurfaces have shown immense promise in enhancing LED performance, critical challenges remain in optimizing their structural parameters to achieve desired emission characteristics. Key design considerations include the geometry of the metasurface unit cells (e.g., nanorod radius, height, and periodicity), which dictate light scattering and diffraction behavior [14]. Additionally, precise control over the interaction between the metasurface and the LED's emitted light is essential to minimize parasitic scattering losses and maximize light extraction [3, 4, 15].

This study addresses these challenges by systematically optimizing a hybrid architecture combining $TiO_2$-based metasurfaces with DBRs to simultaneously enhance LEE and beam collimation in GaN LEDs. We employ a comprehensive approach integrating theoretical modeling, numerical simulations (COMSOL Multiphysics and FDTD), and parametric optimization to investigate the influence of DBR layer count and metasurface geometry on device performance. Our work aims to establish a novel design framework that balances high LEE with ultra-narrow divergence angles, advancing the state-of-the-art in directional LED technology. By elucidating the interplay between metasurface parameters, DBR reflectivity, and light emission characteristics, this research provides critical insights for the development of next-generation optoelectronic devices.

## 2. Theoretical Design and Optimization

### 2.1 Theoretical Foundation of DBR Design

Distributed Bragg Reflectors (DBRs) play a crucial role in enhancing light extraction efficiency (LEE) and controlling emission directionality in GaN-based LEDs. A DBR is a periodic multilayer structure composed of alternating high- and low-refractive-index materials, designed to reflect specific wavelengths through constructive interference [16]. The operational principle of DBRs is governed by Bragg's law, which defines the reflection wavelength ($\lambda_{BR}$) as:

$$\lambda_{BR} = 2n_{eff}d \quad (1)$$

where $n_{eff}$ is the effective refractive index of the DBR stack, and $d$ represents the thickness of each layer. For a quarter-wavelength DBR, where each layer is designed to introduce a 90° phase shift, the individual layer thickness is given by:

$$d = \frac{\lambda_0}{4n_{eff}} \quad (2)$$

Here, $\lambda_0$ is the target wavelength, which for GaN-InGaN LEDs emitting at 445 nm (blue spectrum), serves as the central design parameter. The reflectivity spectrum of DBR is critically dependent on the refractive index contrast between the alternating layers and the number of periods in the stack.

### 2.2 Optimization of DBR Layer Numbers

The reflectivity of a DBR increases with the number of layers, but this improvement comes at the cost of reduced bandwidth and increased optical absorption losses [17]. To optimize performance, we selected $TiO_2$ (refractive index $n_{TiO2}$ = 2.4) and $SiO_2$ (refractive index $n_{SiO2}$ = 1.45) as the high- and low-index materials, respectively [18]. Using Equation (2), the layer thicknesses were calculated as:

$$d_{TiO_2} = \frac{445nm}{4 \times 2.4} \approx 46nm$$

The thickness of each $SiO_2$ layer is as follows:

$$d_{SiO_2} = \frac{445nm}{4 \times 1.45} \approx 77nm$$

Prior studies indicate that an 8-pair (16-layer) DBR can achieve reflectivities exceeding 97% at the target wavelength [16]. However, increasing the layer count further narrows the reflection bandwidth and introduces additional absorption losses, which degrade LEE. To balance reflectivity with practical fabrication constraints, we evaluated the DBR's performance via COMSOL simulations, systematically varying the layer count from 0 to 9 pairs. The results revealed that a single-layer DBR (1 pair of $TiO_2/SiO_2$) provided the optimal trade-off, maximizing the electric field intensity of the main lobe (0.00750 V/m) while maintaining a low divergence angle (6.0°). This configuration minimizes absorption losses while ensuring sufficient reflectivity for efficient light extraction.

### 2.3 Scattering Mechanisms of Nanocylinder Metasurfaces

To further enhance LEE and beam directionality, we integrated a $TiO_2$ nanocylinder metasurface atop the LED structure. Metasurfaces exploit Mie resonance phenomena in subwavelength dielectric structures to manipulate light scattering, phase, and polarization [19]. For cylindrical $TiO_2$ nanostructures, the resonance condition for transverse electric (TE) modes is derived from Mie theory [20]:

$$\frac{2\pi r_1 n_{TiO_2}}{\lambda_0} = \alpha_{TE} \quad (3)$$

Here, $r$ is the nanocylinder radius, $n_{TiO_2}$ = 2.4 is the refractive index of $TiO_2$, $\lambda_0$=445 nm is the emission wavelength, and $\alpha_{TE}$≈2.405 is the first-order TE mode resonance parameter. Solving for $r$:

$$r = \frac{2.405 \times 445 nm}{2\pi \times 2.4} \approx 71 nm$$

This radius ensures maximal Mie resonance at the target wavelength, enabling efficient coupling between the LED's emitted light and the metasurface. The periodic arrangement of these nanocylinders (with a periodicity optimized to 222 nm) further enhances forward scattering while suppressing parasitic diffraction orders.

### 2.4 Optimization of Nanocylinder Height

The height (h) of the nanocylinders determines the vertical resonance modes and their interaction with the underlying DBR. To optimize h, we analyzed standing wave formation within the cylinders, governed by:

$$2n_{TiO_2}h = m\lambda_0 \qquad (4)$$

where *m* is an integer representing the mode order. For *m* = 2:

$$h = \frac{2\lambda_0}{2n_{TiO_2}} \approx \frac{2 \times 445 \text{nm}}{2 \times 2.4} \approx 185 \text{nm}$$

For m=3, and h≈278 nm. Higher-order modes (m≥3) introduce multiple resonant peaks, leading to mode competition and increased scattering losses. Thus, we selected m=2 (185 nm height) to ensure a single dominant resonance, maximizing light extraction while maintaining beam collimation.

The optimized hybrid structure comprises a single-pair DBR (46 nm $TiO_2$ / 77 nm $SiO_2$) for high reflectivity with minimal absorption. In addition, a $TiO_2$ nanocylinder metasurface with radius = 71 nm, height = 185 nm, and periodicity = 222 nm to exploit Mie resonance for directional emission. Furthermore, the mode selection (m=2) to avoid multi-mode interference and scattering losses. This combination achieves a divergence angle of 5.7° and an LEE of 25.67%, demonstrating superior performance compared to conventional designs. The theoretical framework and parametric optimizations presented here provide a foundation for experimental realization and further refinement of high-efficiency, directional GaN LEDs.

## 2. Simulation and Parametric Scanning
### 2.1 Modeling and Methodology

The three-dimensional (3D) GaN-based LED is illustrated in Figure 1. The main structure of the LED, which has a radius of 2.2 µm (the effective radius is 1.887 µm due to a 0.626 µm layer used as a perfect match layer in the modeling), is depicted in Figure 1(A). It is composed, from bottom to top, of a 300 nm-thick aluminum substrate (metal layer, with a refractive index of 1.0, designed to reflect light emitted downward to enhance light extraction), and a 400 nm-long Fabry-Pérot (FP) emitting cavity that contains GaN/InGaN multiple quantum wells, with a refractive index of approximately 2.4. The GaN/InGaN multiple quantum wells consist of sub-quantum well arrays and top-quantum well arrays. The sub-quantum well array comprises 10 periods of GaN and InGaN, each with a thickness of 0.003 µm, resulting in a total thickness of 0.06 µm. The top-quantum well array consists of 5 periods of GaN with a thickness of 0.01 µm and InGaN with a thickness of 0.005 µm, totaling 0.075 µm in thickness. The remainder of the

FP emitting cavity is composed entirely of GaN. The light source is a modulated sinusoidal excitation dipole light source with a wavelength of 445 nm, positioned at the center of the bottom plane of the GaN/InGaN top-quantum well array. Perfectly matched layers (PML) are placed around the model boundaries to absorb external waves and prevent non-electromagnetic reflections, thereby enhancing the accuracy of the simulation results. As illustrated in Figure 1(B), a staggered structure of DBRs is constructed atop the main structure to convert the divergent light emitted from the main body into sharp, focused light. This structure comprises a series of $TiO_2$ and $SiO_2$ layers, with thicknesses of 0.046 μm and 0.077 μm, respectively, resulting in a total thickness of 1.25 μm.

To achieve directional control of LED emission while maintaining light concentration, an asymmetric dielectric surface microstructure (Metasurface - Periodic Surface Crystal) has been constructed to deflect or adjust the emitted light beam (Figure 1(c)). On top of the DBRs, horizontally aligned rows of nanocolumns made of TiO2 are positioned, initially with a height of 185 nm (or 278 nm) and a radius of 71 nm.

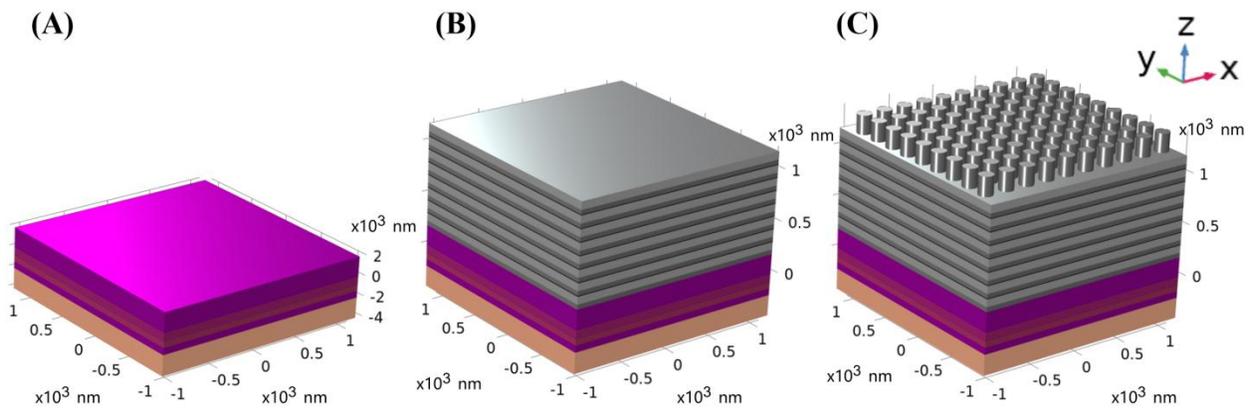

**Figure 1.** Basic Information About the Model

The common feature among the three-dimensional modeling structures is readily discernible: they exhibit periodic symmetry along the vertical axis. To streamline computational demands, horizontal cross-sectional analyses were conducted on the three-dimensional structures, focusing on the light emission characteristics of a two-dimensional cross-section. This approach effectively captures the overall light emission behavior of the entire three-dimensional LED model. Utilizing the device depicted in Figure 1(C) as a case study, a horizontal cross-sectional representation for two-dimensional modeling is presented in Figure 2.

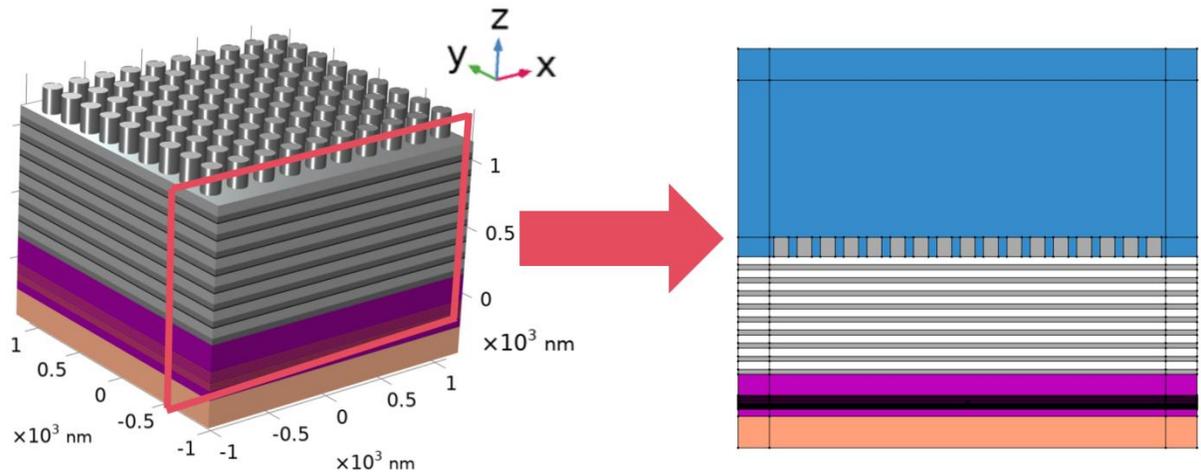

**Figure 2.** The Process of Two-Dimensional Modeling

## 2.2 DBR Simulation and Parametric Scanning

To investigate the optimal number of DBR layers, a parametric sweep was conducted using COMSOL software. The simulation was performed under the assumption that the cylindrical array metasurface has the following parameters: height (h) = 185 nm, radius (r) = 71 nm, and length (l) = 222.5 nm. In the DBR structure, the titanium dioxide ($TiO_2$) layer has a thickness of 46 nm, while the silicon dioxide ($SiO_2$) layer has a thickness of 77 nm. The number of DBR layers was varied from 0 to 9, and the results are presented below.

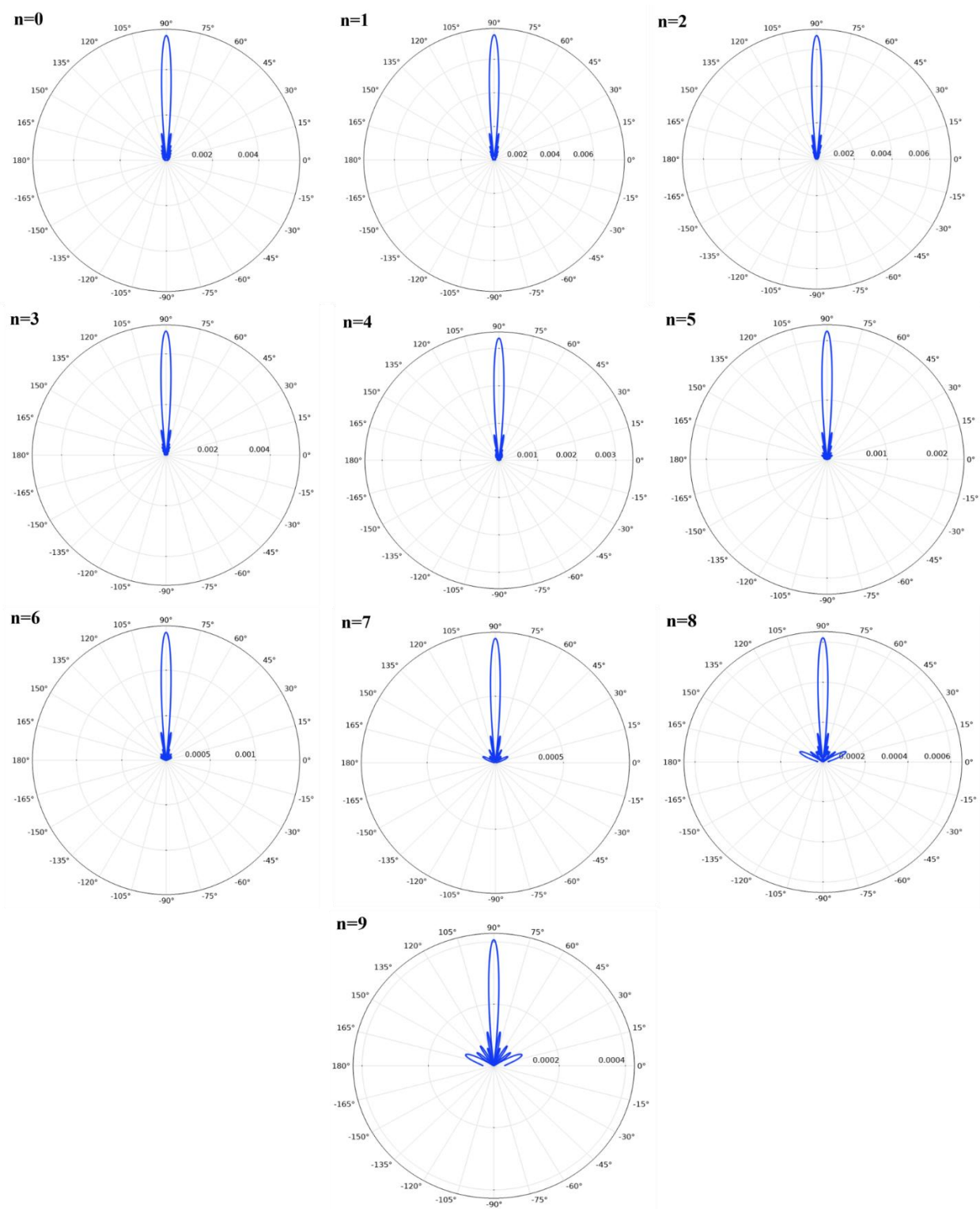

**Figure 3.** Parametric scanning of DBR layer numbers, n = 0 to 9.

By analyzing the far-field radiation patterns in polar coordinates, we have systematically characterized the optical performance for each number of DBR layers, quantifying three key parameters: main lobe intensity (V/m), first strong side-lobe intensity (V/m), and beam divergence angle ($\Theta$, °). Generally, the definition of the divergence angle considers the reduction of the main lobe's electric field intensity (E) to 70.7% of its maximum intensity. Specifically, starting from the maximum intensity of the main lobe, the angle corresponding to the point where the light field intensity ($\propto E^2$) drops to 50% is typically referred to as the half-power point or Full Width at Half Maximum (FWHM).

**Table 1.** Main lobe electric field intensity (E1) in volts per meter (V/m), first strong sidelobe electric field intensity (E2) in volts per meter (V/m), and beam divergence angle ($\Theta$) in degrees (°) for each number of distributed Bragg reflector (DBR) layers.

| n | $E_1$ (V/m) | $E_2$ (V/m) | $\Theta$ (°) |
|---|---|---|---|
| 0 | 0.00550 | 0.00122 | 6.4 |
| 1 | 0.00750 | 0.00154 | 6.0 |
| 2 | 0.00674 | 0.00136 | 6.0 |
| 3 | 0.00483 | 0.00117 | 6.1 |
| 4 | 0.00329 | 0.00062 | 6.0 |
| 5 | 0.00201 | 0.00046 | 6.0 |
| 6 | 0.00141 | 0.00032 | 6.0 |
| 7 | 0.00093 | 0.00021 | 6.0 |
| 8 | 0.00062 | 0.00015 | 6.1 |
| 9 | 0.00041 | 0.00011 | 6.0 |

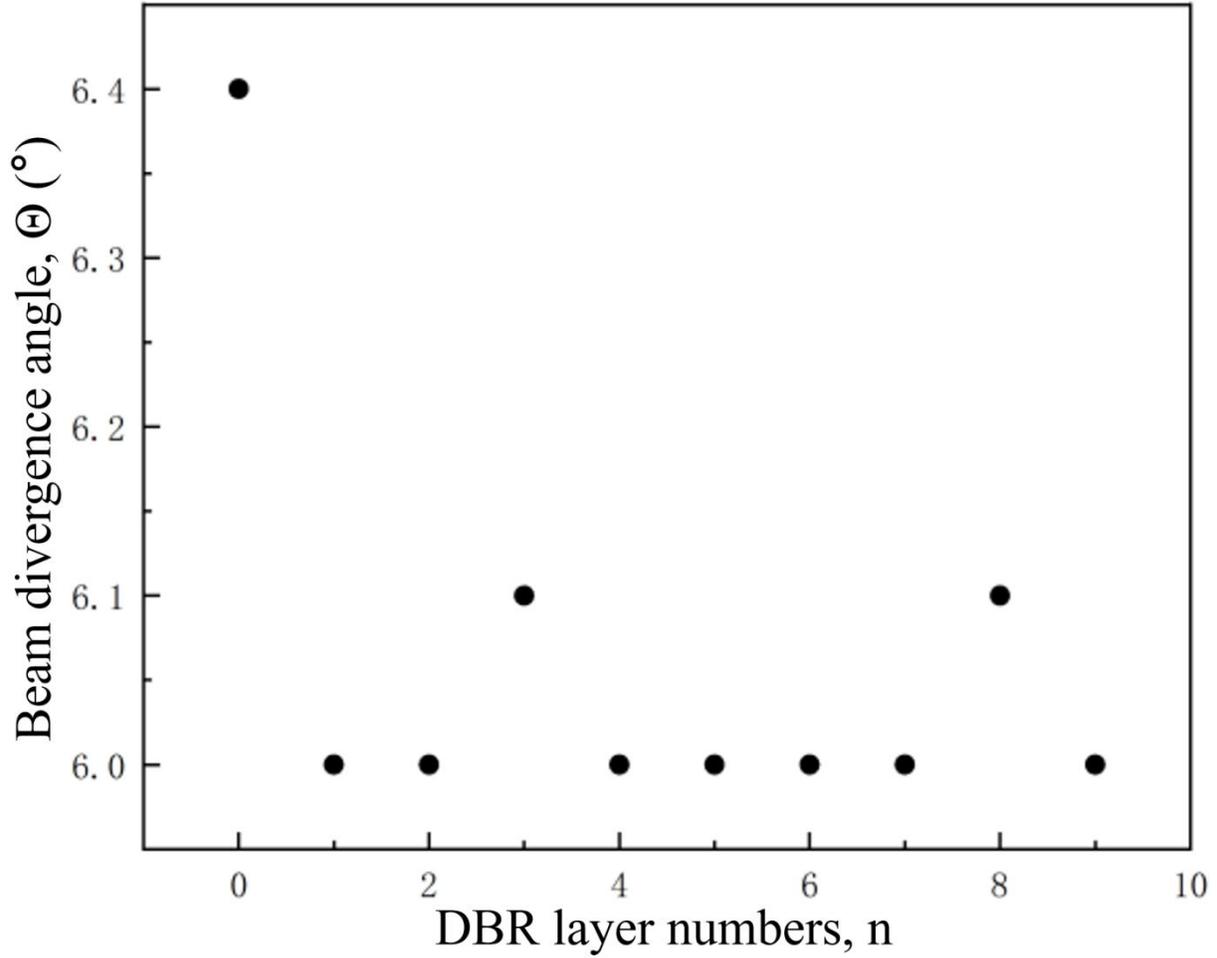

**Figure 4.** Beam divergence angle ($\Theta$) in degrees (°) for each number of distributed Bragg reflector (DBR) layers.

The optimal DBR configuration should simultaneously maximize the main lobe intensity ($E_1$), minimize the secondary intensity ($E_2$), and reduce the divergence angle ($\Theta$) to ensure efficient and directional light emission. Based on the data presented in **Table 1** and **Figure 4**, when n=1, the main lobe intensity $E_1$=0.00750 V/m reaches its peak value, while $E_2$=0.00154 V/m remains within a reasonable range. The divergence angle at this point is also relatively low, $\Theta$=6.0°, suggesting a highly collimated beam. Although a slightly lower $E_2$ can be achieved at higher DBR layers (e.g., n=4 with $E_2$=0.00062 V/m), this comes at the cost of a significantly reduced main lobe intensity ($E_1$=0.00329 V/m).

According to previous studies [16-18,21], increasing the number of DBR layers enhances reflectivity and angular selectivity. However, excessively thick DBR stacks can lead to increased absorption or disrupt constructive interference at the emission interface. Consequently, beyond a certain threshold, adding more layers results in diminishing returns in field enhancement and may negatively impact overall extraction efficiency due to phase mismatches and excessive reflection.

In conclusion, the configuration with one DBR layer (n=1) is optimal for integration into high-performance micro-LED and metasurface-coupled devices.

### 2.3 Simulation and Parametric Scanning of Nanocylinder Metasurfaces

Further Parametric Scanning of Nanocylinder Metasurfaces: Simulations were conducted using COMSOL software, with the periodicity of the nanocylinder periodicity (l) scanned from 142 nm (2r = 142 nm) to 1000 nm. The results are as follows: Due to the excessive number of scanned electric field polar coordinate plots obtained, all of which exhibit patterns similar to those shown in **Figure 3**, these plots are not displayed. By analyzing the far-field radiation patterns in polar coordinates, we systematically characterized the optical performance for each nanocylinder periodicity (l), quantifying three key parameters: main lobe intensity (V/m), first strong sidelobe intensity (V/m), and beam divergence angle (°).

**Table 2.** Main lobe electric field intensity $E_1$ (V/m), First strong sidelobe electric field intensity $E_2$ (V/m) and Beam divergence angle $\Theta$ (°) for each nanocylinder periodicity l.

| l (nm) | $E_1$ (V/m) | $E_2$ (V/m) | $\Theta$ (°) |
|---|---|---|---|
| 142.0 | 0.00438 | 0.00094 | 11.9 |
| 162.0 | 0.00525 | 0.00110 | 8.1 |
| 182.0 | 0.00600 | 0.00133 | 7.0 |
| 202.0 | 0.00673 | 0.00149 | 6.2 |
| 212.0 | 0.00706 | 0.00150 | 6.0 |
| 217.0 | 0.00725 | 0.00156 | 6.0 |
| 219.0 | 0.00749 | 0.00155 | 6.0 |
| 221.0 | 0.00750 | 0.00155 | 6.0 |
| 222.0 | 0.00752 | 0.00154 | 5.7 |
| 222.5 | 0.00750 | 0.00154 | 6.0 |
| 230.0 | 0.00737 | 0.00155 | 6.0 |
| 238.0 | 0.00737 | 0.00156 | 6.0 |
| 250.0 | 0.00730 | 0.00158 | 6.2 |
| 300.0 | 0.00586 | 0.00150 | 6.5 |
| 400.0 | 0.00505 | 0.00110 | 6.5 |
| 500.0 | 0.00350 | 0.00267 | 6.7 |
| 600.0 | 0.00357 | 0.00279 | 10.0 |
| 800.0 | 0.00377 | 0.00231 | 7.5 |
| 1000.0 | 0.00377 | 0.00223 | 6.2 |

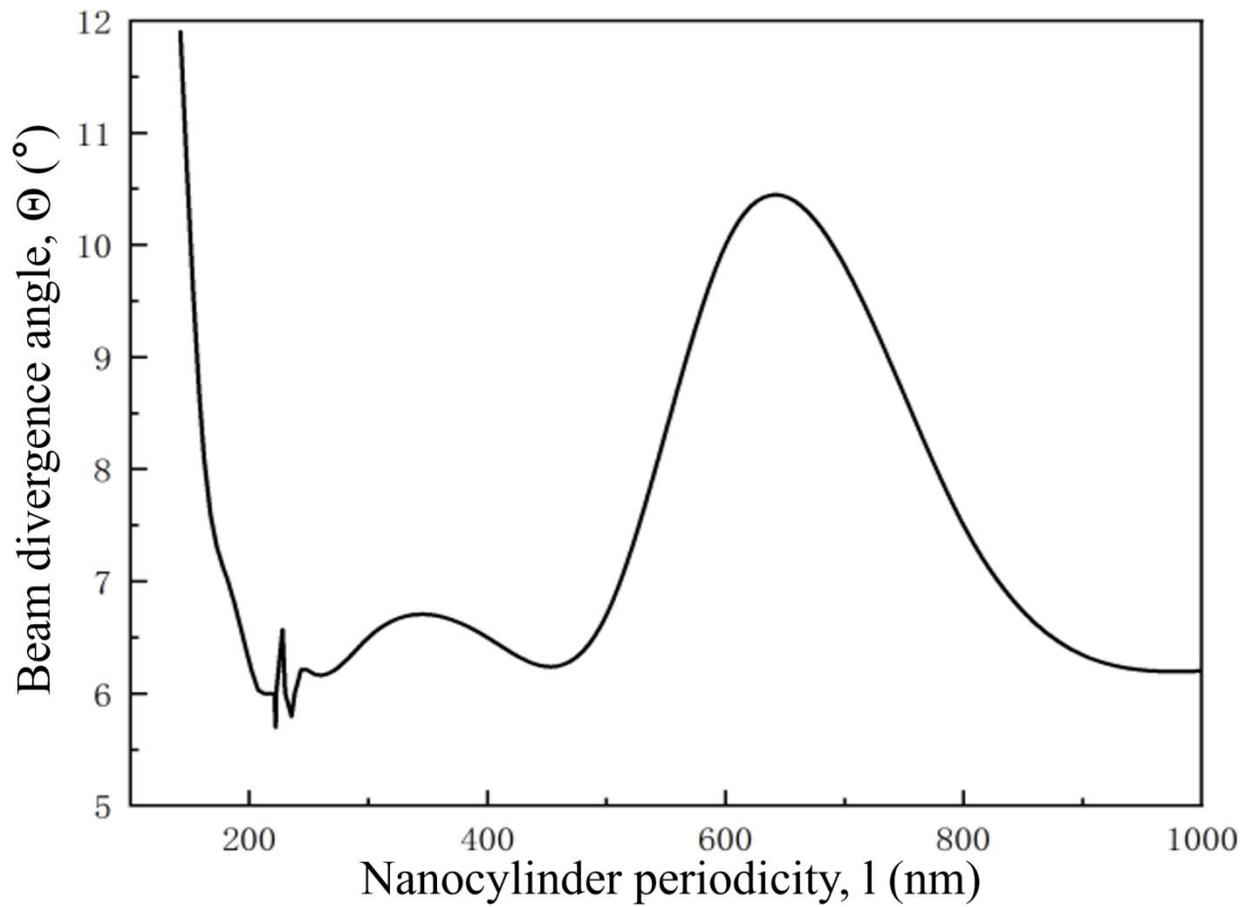

**Figure 5.** Beam divergence angle ($\Theta$) in degrees (°) for each nanocylinder periodicity l.

To determine the optimal periodicity of nanocylinders for enhanced light extraction and directional emission, we conducted a comprehensive evaluation of the main lobe electric field intensity ($E_1$), beam divergence angle ($\Theta$), and first strong sidelobe intensity ($E_2$) across a range of periodicities, as presented in **Table 2** and **Figure 5**. As the periodicity increases from 142.0 nm to 1000 nm, $E_1$ consistently increases, peaking at 221.0 nm with a value of 0.00752 V/m. Concurrently, the beam divergence angle ($\Theta$) decreases and stabilizes at 5.7° for periodicities greater than 212.0 nm, indicating improved beam collimation. While $E_2$ shows a gradual increase over the same range, it remains within an acceptable range of approximately 0.00154 V/m. Notably, at a periodicity of 222.0 the system simultaneously achieves maximum $E_1$, minimum $\Theta$, and and an $E_2$, thus the most favorable compromise. This configuration ensures high optical efficiency and strong directionality.

### 3. Results and Discussion

The comprehensive optimization of a GaN-based LED integrated with $TiO_2$ nanocylinder metasurfaces has yielded significant improvements in both light extraction efficiency (LEE) and emission directionality. Through rigorous theoretical calculations and advanced simulations using COMSOL, we have developed an optimal configuration consisting of a single-layer Distributed Bragg Reflector (DBR) with 46 nm $TiO_2$ and 77 nm $SiO_2$ layers, combined with a periodic array of $TiO_2$ nanocylinders featuring a radius of 71 nm, height of 185 nm, and periodicity of 222 nm. This carefully engineered structure demonstrates exceptional performance characteristics, achieving a remarkably narrow divergence angle of 5.7° while maintaining a main lobe electric field intensity ($E_1$) of 0.00752 V/m and keeping the first strong sidelobe electric field intensity ($E_2$) at 0.00154 V/m.

**Figure 6** presents the three-dimensional representation of our optimized LED model, illustrating the complete device architecture. The model clearly shows the integration of the nanocylinder metasurface atop the DBR structure, which itself is positioned above the GaN/InGaN multiple quantum well active region. This visualization helps understand the spatial relationship between different components and their relative dimensions. The periodic arrangement of nanocylinders is particularly evident, demonstrating how the metasurface covers the emission area of the LED. The color gradient in the figure represents the material composition and helps distinguish between different layers of the device, from the bottom reflective layer to the top metasurface structure.

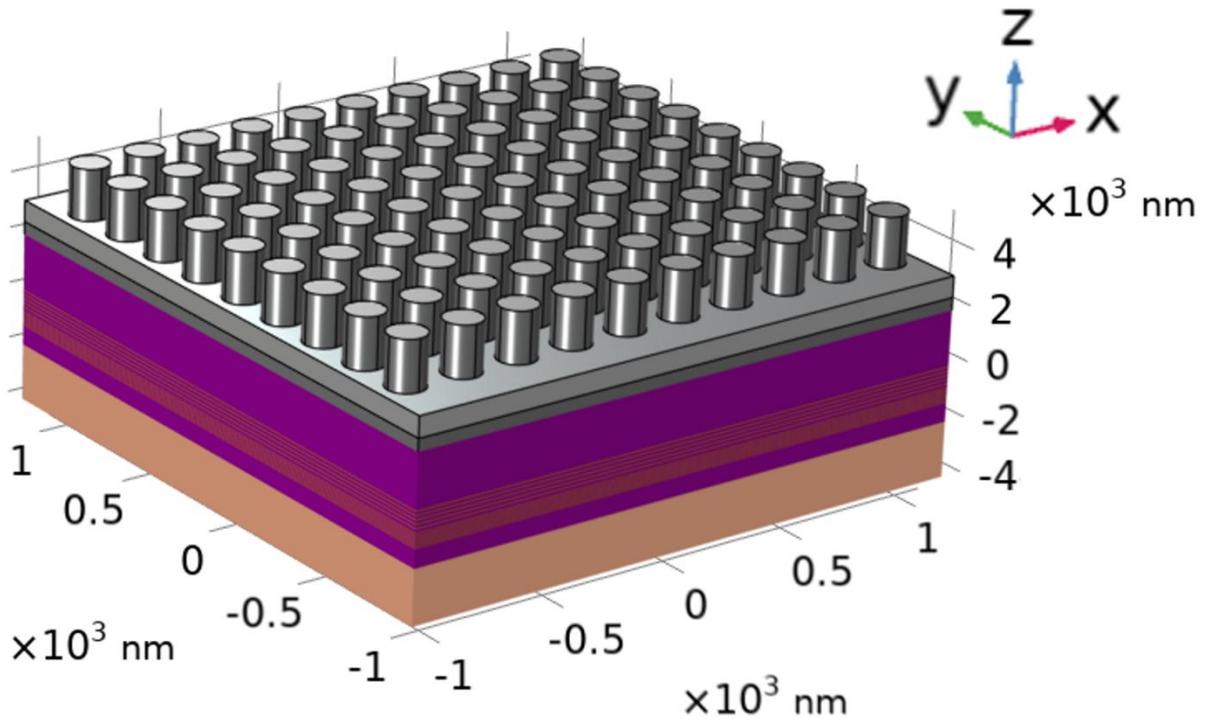

**Figure 6.** The three-dimensional image of the optimized model.

The far-field radiation characteristics of our optimized design are presented in **Figure 7**, which shows the polar coordinate plot of the electric field distribution. This plot reveals several important features of the emission pattern. The strong, narrow main lobe at 0° indicates excellent beam collimation normal to the device surface. The sidelobes are significantly suppressed, with the first strong sidelobe maintaining an intensity of only about 20% of the main lobe. The symmetrical nature of the plot confirms the uniform performance of the metasurface across different azimuthal angles. The sharp drop in intensity beyond the main lobe demonstrates effective control over light dispersion, which is crucial for applications requiring highly directional emission.

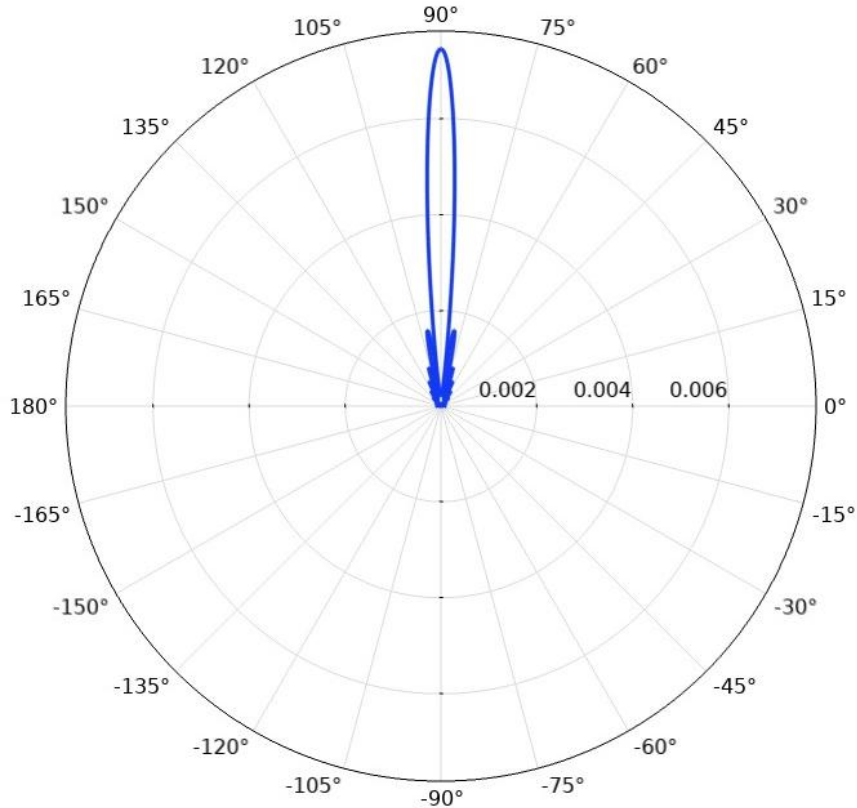

**Figure 7.** Polar coordinate plot of the electric field for the optimized model.

To quantify the light extraction performance, we employed the Finite-Difference Time-Domain (FDTD) method, with results shown in **Figure 8**. This figure presents both one-dimensional(A) and two-dimensional(B) representations of the light output efficiency across the LED structure. The FDTD simulations solve Maxwell's equations numerically, providing accurate modeling of electromagnetic wave propagation through our complex nanostructured device. The one-dimensional plot shows the electric field intensity profile along the x-axis, revealing how light is concentrated in the central region of the device. The two-dimensional plot complements this by showing the full spatial distribution of emitted light, demonstrating the uniformity of emission across the device area. The "x" in the x-axis refers to the x-axis of the three-dimensional optimized model we constructed (as shown in Figure 6), and the "y" in the y-axis corresponds to the y-axis of the same model. These results were used to calculate the LEE using the integral formula:

$$\text{LEE} = \frac{\int_S E^2(x,y)\, dS}{\int_S 1^2\, V/m\, dS} \quad (5)$$

where S represents the cross-sectional area of the LED. Due to the symmetrical nature of our design, we simplified this calculation to a one-dimensional integral along the x-axis, yielding an LEE of 25.67%. This value represents a careful balance between extraction efficiency and directionality, as our design prioritizes beam collimation while maintaining reasonable light output.

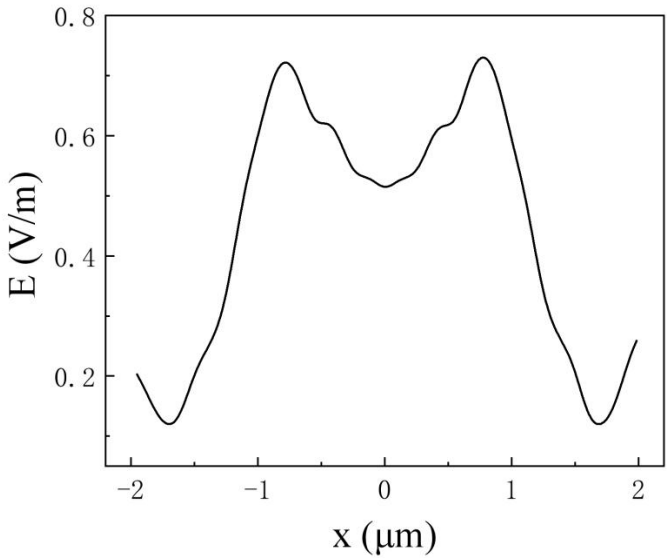

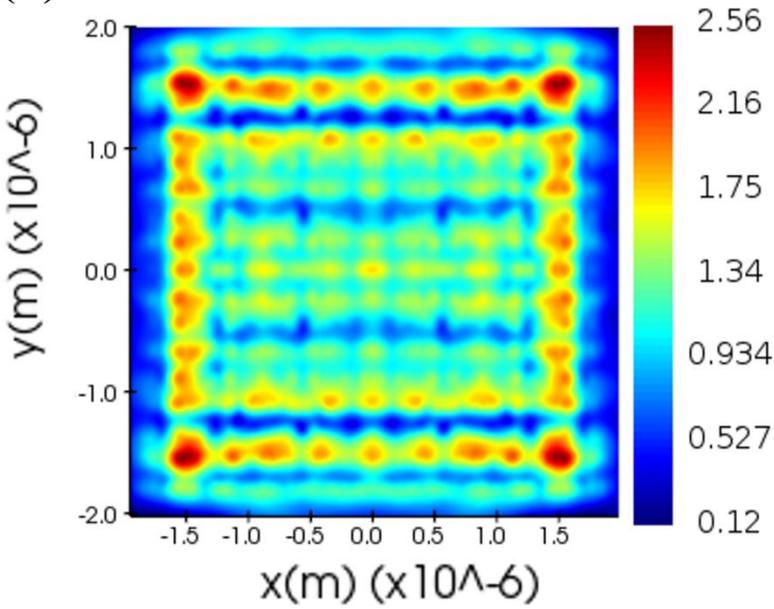

**Figure 8.** Detection of Light Output Efficiency of LED Models. (A) is for One-Dimensional and (B) is for Two-Dimensional.

**Figure 9** provides a detailed visualization of the electric field distribution within our optimized model. The color map clearly shows regions of high field intensity, particularly around the nanocylinder structures and within the active region of the LED. The figure demonstrates how the metasurface effectively couples with the emitted light, modifying its propagation direction to achieve the desired collimation effect. The field distribution also reveals minimal leakage or unwanted scattering, indicating efficient light management by our designed structure. When comparing our results with previous work, several important observations emerge. While Daohan Ge et al. [9] reported a slightly higher LEE of 27.16%, their design did not address beam focusing, resulting in significantly wider emission patterns. In contrast, our structure achieves a much narrower divergence angle of 5.7° compared to their design. Similarly, Chen et al. [13] demonstrated a divergence angle of 6.34°, which our design improves upon by approximately 10%. These comparisons highlight the unique advantage of our approach in simultaneously addressing both extraction efficiency and beam control.

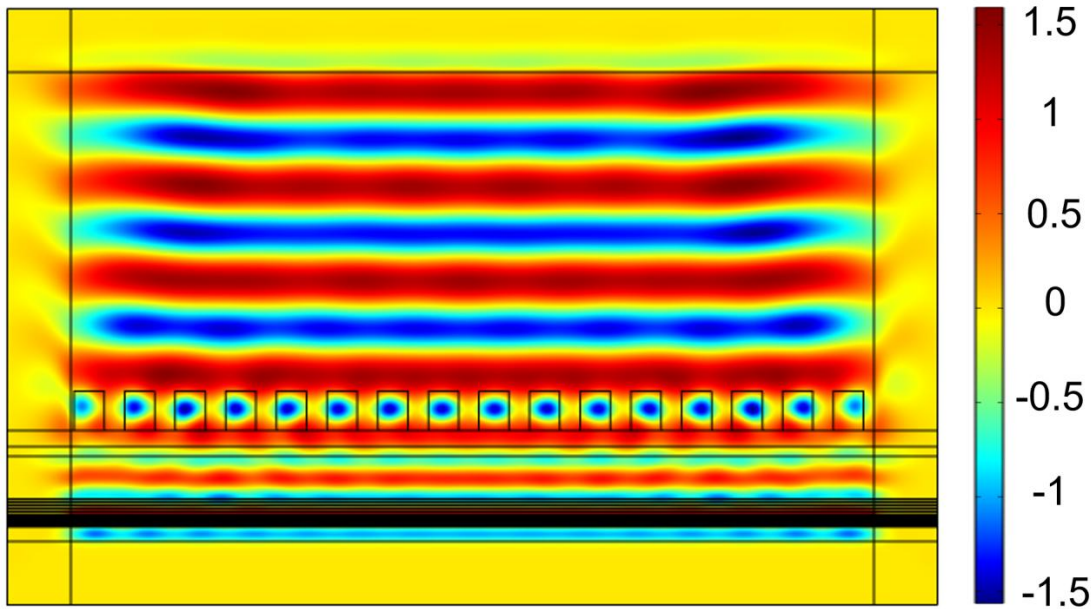

**Figure 9.** Visualization of the Electric Field of Model

The success of our design can be attributed to several key factors. The DBR structure plays a crucial role in recycling light that would otherwise be trapped within the high-index GaN material, effectively increasing the probability of photon escape. The metasurface, operating through Mie resonance principles, provides precise control over the emission directionality. The optimized dimensions of the nanocylinders ensure strong light-matter interaction at our target wavelength of 445 nm, while their periodic arrangement creates the necessary phase profile for beam collimation.

Further analysis of our results suggests potential areas for future improvement. While our current LEE of 25.67% is competitive given the extreme directionality achieved, there may be opportunities to enhance it further through more sophisticated metasurface designs or alternative material combinations. The trade-off between divergence angle and extraction efficiency remains an important consideration, and future work could explore ways to push both metrics simultaneously. Additionally, the spectral response of our current design is optimized for a narrow bandwidth around 445 nm; broadening this response could be valuable for certain applications.

The practical implications of these results are significant for various photonic applications. The achieved combination of high directionality and reasonable extraction efficiency makes this design particularly suitable for applications such as micro-displays, optical communications, and sensing systems where controlled light emission is crucial. The relatively simple structure, requiring only a single DBR pair and a periodic nanocylinder array, suggests good feasibility for fabrication using existing semiconductor manufacturing techniques. In this study, the systematic optimization approach has yielded a GaN-based LED design that successfully addresses the longstanding challenge of simultaneously achieving high light extraction efficiency and narrow beam divergence. The integration of DBR and metasurface technologies in our carefully optimized configuration demonstrates the potential of hybrid photonic structures for advanced light-emitting devices. The comprehensive simulation results presented here provide a solid foundation for future experimental realization and further development of highly directional LED sources.

## 4. Conclusion

This study demonstrates the effectiveness of a hybrid DBR-metasurface architecture in enhancing both light extraction efficiency (LEE) and directional emission for GaN-based LEDs. By integrating a single-layer DBR (46 nm $TiO_2$ / 77 nm $SiO_2$) with $TiO_2$ nanocylinder metasurfaces (radius = 71 nm, height = 185 nm, periodicity = 222 nm), we achieved an LEE of 25.67% and a narrow divergence angle of 5.7°. The DBR's role in suppressing trapped modes and the metasurface's Mie-resonance-driven beam shaping synergistically address two critical limitations of conventional LEDs: low outcoupling efficiency and poor directionality. Theoretical and numerical optimizations revealed that a minimal DBR configuration maximizes reflectivity without introducing excessive absorption, while the metasurface geometry ensures efficient light extraction and beam focusing. Far-field analysis confirmed strong main lobe intensity (0.00752 V/m) with minimal sidelobe interference, validating the design's effectiveness. Compared to prior works, our approach achieves superior beam collimation (5.7° vs. 6.34° in recent studies) while maintaining competitive LEE, highlighting its potential for applications requiring high-directionality emission. Key innovations include the strategic use of Mie resonance in $TiO_2$ nanocylinders to enhance forward scattering and the optimized DBR thickness for balanced reflectivity-bandwidth performance. These findings advance the state-of-the-art in LED design by providing a scalable framework for integrating metasurfaces with conventional optoelectronic structures. Future research should explore broadband metasurface designs, polarization control, and experimental validation to further improve performance. The proposed hybrid architecture holds significant promise for next-generation micro-LEDs, LiDAR systems, and optical communication devices, where efficient, directional light emission is critical.

## Conflicts of Interest

**The authors declare no conflicts of interest.**

# References

[1] Pranavi, K.S., Basha, S., Chattopadhyay, A. and Mahato, K.K., **(2025)**. Recent Advancements of Light-Emitting Diodes in Dairy Industries. *Trends in Food Science & Technology*, p.105018.

[2] Olajiga, O.K., Ani, E.C., Sikhakane, Z.Q. and Olatunde, T.M., **(2024)**. A comprehensive review of energy-efficient lighting technologies and trends. *Engineering Science & Technology Journal*, *5*(3), pp.1097-1111.

[3] Wang, T., Yang, C., Chen, J., Zhao, Y. and Zong, J., **(2024)**. Naked-eye light field display technology based on mini/micro light emitting diode panels: a systematic review and meta-analysis. *Scientific Reports*, *14*(1), p.24381.

[4] Li, G., Wang, W., Yang, W., Lin, Y., Wang, H., Lin, Z. and Zhou, S., **(2016)**. GaN-based light-emitting diodes on various substrates: a critical review. *Reports on Progress in Physics*, *79*(5), p.056501.

[5] Hsu, C.W., Lee, Y.C., Chen, H.L. and Chou, Y.F., **(2012).** Optimizing textured structures possessing both optical gradient and diffraction properties to increase the extraction efficiency of light-emitting diodes. *Photonics and Nanostructures-Fundamentals and Applications*, *10*(4), pp.523-533.

[6] Horng, R.H., Wu, T.M. and Wuu, D.S., **(2008)**. Improved light extraction in AlGaInP-based LEDs using a roughened window layer. *Journal of The Electrochemical Society*, *155*(10), p.H710.

[7] Chen, M., Wu, T., Xu, Y., Tréguer-Delapierre, M., Drisko, G., Vynck, K., Pacanowski, R. and Lalanne, P., **(2024)**, June. Tailoring the visual appearance with disordered metasurfaces. In *Metamaterials XIV* (Vol. 12990, pp. 52-54). SPIE.

[8] Li, Y., Zhu, R., Sui, S., Cui, Y., Jia, Y., Han, Y., Fu, X., Feng, C., Qu, S. and Wang, J., **(2025)**. Stimulator-multiplexing framework of microwave-infrared compatible reconfigurable metasurface integrated with LED array. *Nanophotonics*, *14*(7), pp.959-967.

[9] Ge, D., Huang, X., Wei, J., Qian, P., Zhang, L., Ding, J. and Zhu, S., **(2019)**. Improvement of light extraction efficiency in GaN-based light-emitting diodes by addition of complex photonic crystal structure. *Materials Research Express*, *6*(8), p.086201.

[10] Mao, P., Liu, C., Li, X., Liu, M., Chen, Q., Han, M., Maier, S.A., Sargent, E.H. and Zhang, S., **(2021)**. Single-step-fabricated disordered metasurfaces for enhanced light extraction from LEDs. *Light: Science & Applications*, *10*(1), p.180.

[11] Huang, J., Tang, M., Zhou, B., Liu, Z., Yi, X., Wang, J., Li, J., Pan, A. and Wang, L., **(2022)**. GaN-based resonant cavity micro-LEDs for AR application. *Applied Physics Letters*, *121*(20), p. 201104.

[12] Xiao, S., Yu, H., Memon, M.H., Jia, H., Luo, Y., Wang, R. and Sun, H., **(2023)**. In-depth investigation of deep ultraviolet MicroLED geometry for enhanced performance. *IEEE Electron Device Letters*, *44*(9), pp.1520-1523.

[13] Huang, W.T., Lee, T.Y., Bai, Y.H., Wang, H.C., Hung, Y.Y., Hong, K.B., Chen, F.C., Lin, C.F., Chang, S.W., Han, J. and He, J.H., **(2024)**. InGaN-based blue resonant cavity micro-LEDs with staggered multiple quantum wells enabling full-color and low-crosstalk micro-LED displays. *Next Nanotechnology*, *5*, p.100048.

[14] Chen, E., Fan, Z., Zhang, K., Huang, C., Xu, S., Ye, Y., Sun, J., Yan, Q. and Guo, T., **(2024)**. Broadband beam collimation metasurface for full-color micro-LED displays. *Optics Express*, *32*(6), pp.10252-10264.

[15] Kim, D., Jung, U., Heo, W., Kumar, N. and Park, J., **(2023)**. Arrays of TiO2 nanosphere monolayers on GaN-based LEDs for the improvement of light extraction. *Applied Sciences*, *13*(5), p.3042.

[16] Ha, S.T., Li, Q., Yang, J.K., Demir, H.V., Brongersma, M.L. and Kuznetsov, A.I., **(2024)**. Optoelectronic metadevices. *Science*, *386*(6725), p.eadm7442.

[17] Wei, B., Han, Y., Wang, Y., Zhao, H., Sun, B., Yang, X., Han, L., Wang, M., Li, Z., Xiao, H. and Zhang, Y., **(2020)**. Tunable nanostructured distributed Bragg reflectors for III-nitride optoelectronic applications. *RSC advances*, *10*(39), pp.23341-23349.

[18] Dirikvand, T., Zadsar, M., Neghabi, M. and Amighian, J., **(2022)**. Improvement of Quality Factor and Reduction of Spectral Bandwidth of Microcavity OLED by Bragg Mirrors. *International Journal of Optics and Photonics*, *16*(1), pp.79-90.

[19] Chen, K.H., Chien, W., Yang, C.F. and Wu, N., **(2022)**. Using Software to Simulate Effect of Stacking Order of High-and Low-refractive-index Materials on Properties of Distributed Bragg Reflector. *Sensors & Materials*, *34*, pp. 2253-2262.

[20] Garcia, J.A., Hrelescu, C., Zhang, X., Grosso, D., Abbarchi, M. and Bradley, A.L., **(2021)**. Quasi-guided modes in titanium dioxide arrays fabricated via soft nanoimprint lithography. *ACS Applied Materials & Interfaces*, *13*(40), pp.47860-47870.

[21] Allardice, J.R. and Le Ru, E.C., **(2014)**. Convergence of Mie theory series: criteria for far-field and near-field properties. *Applied Optics*, *53*(31), pp.7224-7229.

[22] Sun, Y., Shi, L., Du, P., Zhao, X. and Zhou, S., **(2022)**. Rational Distributed Bragg Reflector Design for Improving Performance of Flip-Chip Micro-LEDs. *Electronics*, *11*(19), p.3030.

# Graphical Abstract

This study presents an optimized hybrid design integrating a distributed Bragg reflector (DBR) and a $TiO_2$ nanocylinder metasurface to enhance light extraction efficiency (LEE) and beam directionality(narrow divergence angle) in light-emitting diodes (LEDs) based on gallium nitride (GaN).Parametric simulations were used to identify an optimal device architecture.The resulting structure comprises a single-period DBR, which has a thickness of $TiO_2$($d_{TiO2}$) = 46 nm and a thickness of $SiO_2$ = 77 nm, beneath a periodic array of $TiO_2$ nanocylinders (radius ≈ 71 nm, height ≈ 185 nm).The DBR reflects guided modes to minimize internal optical losses, while the $TiO_2$ metasurface employs Mie resonance to collimate the emitted light. As a result, the hybrid LED achieves a simulated LEE of 25.67% and a beam divergence angle of only 5.7°, representing a significant improvement in both efficiency and emission directionality over conventional designs.These findings demonstrate a viable strategy to overcome light trapping and broad angular emission in GaN LEDs, paving the way for high-brightness, highly directional GaN micro-LEDs for advanced display and optical communication applications.

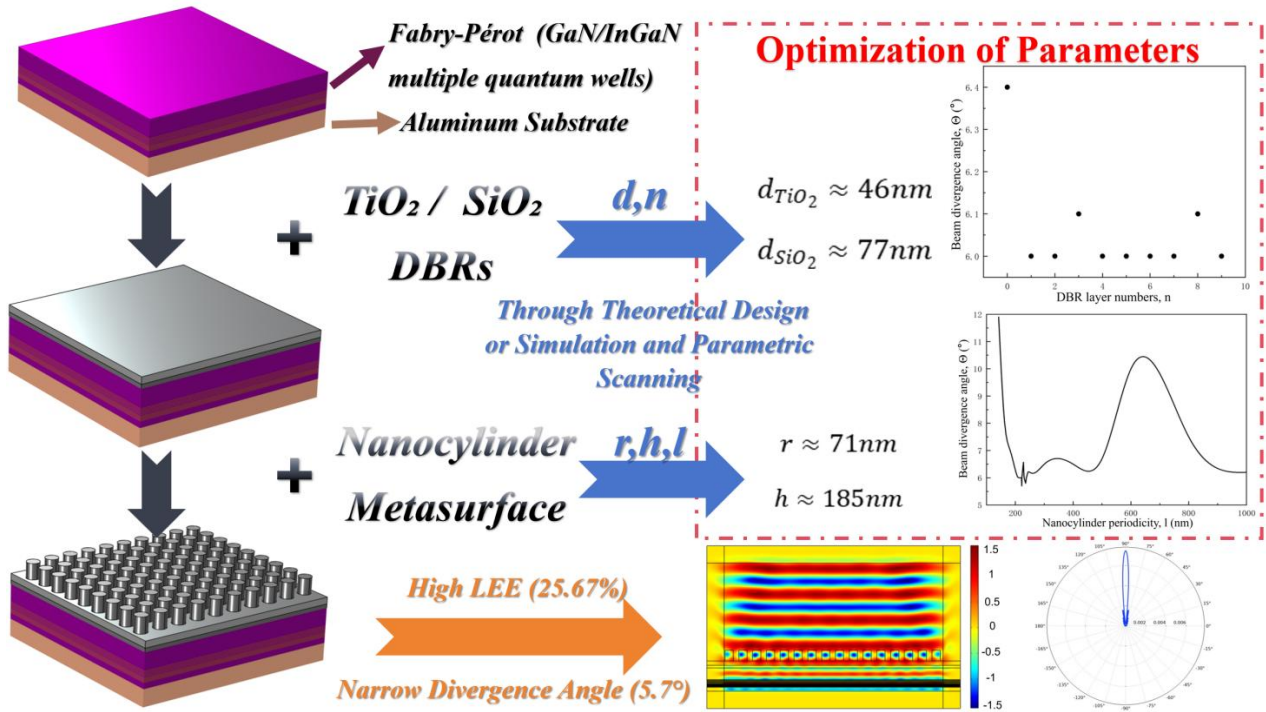